\documentclass{ieeeaccess}
\usepackage{cite}
\usepackage{amsmath,amssymb,amsfonts}
\usepackage{graphicx}
\usepackage{textcomp}
\usepackage{lineno}
\usepackage{epsfig}
\usepackage{chngpage}
\usepackage{float}
\usepackage{amsmath}
\usepackage{graphicx}
\usepackage{array}
\usepackage{diagbox}
\usepackage{epstopdf}
\usepackage{algorithm}
\usepackage{algorithmicx}
\usepackage{algpseudocode}
\usepackage{threeparttable}
\usepackage{subfigure}
\usepackage{multirow}
\usepackage{longtable}
\usepackage[table,xcdraw]{xcolor}
\usepackage{url}
\usepackage{booktabs}
\usepackage{multirow}
\usepackage[colorlinks,linkcolor=blue]{hyperref}
\usepackage{algpseudocode}
\usepackage{amsmath}
\usepackage{amssymb}
\usepackage{caption}

\DeclareMathOperator*{\argmin}{arg\,min}
\def\BibTeX{{\rm B\kern-.05em{\sc i\kern-.025em b}\kern-.08em
    T\kern-.1667em\lower.7ex\hbox{E}\kern-.125emX}}
\begin{document}
\history{Date of publication xxxx 00, 0000, date of current version xxxx 00, 0000.}
\doi{10.1109/ACCESS.2020.DOI}

\title{Network Representation Learning: From Traditional Feature Learning to Deep Learning}
\author{\uppercase{Ke Sun}\authorrefmark{1},
\uppercase{Lei Wang}\authorrefmark{1},
\uppercase{Bo Xu}\authorrefmark{1},
\uppercase{Wenhong Zhao}\authorrefmark{2},
\uppercase{Shyh Wei Teng}\authorrefmark{3},
\uppercase{Feng Xia}\authorrefmark{3}, \IEEEmembership{Senior Member,~IEEE}}

\address[1]{Key Lab for Ubiquitous Network and Service Software of Liaoning Province, School of Software, Dalian University of Technology, Dalian 116620, China}
\address[2]{Ultraprecison Machining Center, Zhejiang University of Technology, Hangzhou 310014, China}
\address[3]{School of Engineering, IT and Physical Sciences, Federation University Australia, Ballarat, VIC 3353, Australia}

\tfootnote{This work is partially supported by Zhejiang Provincial Fundamental Public Welfare Research Program under Grant No. LGG18E050025.}

\markboth
{K. Sun \headeretal: Network Representation Learning}
{K. Sun \headeretal: Network Representation Learning}

\corresp{Corresponding author: Feng Xia (e-mail: f.xia@ieee.org).}

\begin{abstract}
Network representation learning (NRL) is an effective graph analytics technique and promotes users to deeply understand the hidden characteristics of graph data. It has been successfully applied in many real-world tasks related to network science, such as social network data processing, biological information processing, and recommender systems. Deep Learning is a powerful tool to learn data features. However, it is non-trivial to generalize deep learning to graph-structured data since it is different from the regular data such as pictures having spatial information and sounds having temporal information. Recently, researchers proposed many deep learning-based methods in the area of NRL. In this survey, we investigate classical NRL from traditional feature learning method to the deep learning-based model, analyze relationships between them, and summarize the latest progress. Finally, we discuss open issues considering NRL and point out the future directions in this field.
\end{abstract}
 

\titlepgskip=-15pt

\maketitle

\section{Introduction}\label{sec:introduction}
\PARstart{R}{epresentation} learning is a new paradigm in the machine learning field aiming at representing information efficiently. For example, in the linguistic domain, word vectors generated from Word2vec framework~\cite{mikolov2013efficient,mikolov2013distributed,morin2005hierarchical} embed semantic information into low dimensional vectors so that machines can better understand the words after word embeddings. In the setting of network representation learning (NRL), a better representation method can make subsequent learning tasks easier. In general, a better representation of a network can preserve the graph topology and cluster similar nodes together in the embedding space. Additionally, the representation learning is beneficial for lots of downstream tasks, e.g., clustering~\cite{fortunato2010community}, node classification~\cite{grover2016node2vec}, and link-prediction~\cite{tang2015line}. They have been widely applied in bioinformatics~\cite{hamilton2017inductive}, linguistics~\cite{mikolov2013efficient}, transport network~\cite{xia2019ranking,wang2020vehicle} and social sciences~\cite{yu2017team,zhang2019judging,kong2019academic,xu2020multivariate}, etc. Many information processing tasks mentioned above depend on how the data are represented. Meanwhile, with the development of big data, which presents Volume, Variety, and Velocity characteristics, a more effective data representation method is required to achieve low cost and tractable computation.

In the past few decades, many traditional feature learning (TFL) algorithms e.g., principal component analysis (PCA)~\cite{wold1987principal}, isometric feature mapping (Isomap)~\cite{tenenbaum2000global}, and local linear embedding (LLE)~\cite{roweis2000nonlinear}, have been proposed for reducing dimensions of data. Instead, NRL focuses on learning the vector representation of a node or a graph. In general, these two kinds of algorithms are then combined to take advantage of both of them.

However, these traditional methods could not effectively extract complex and nonlinear structured relationships of data. It is widely recognized that deep learning~\cite{lecun2015deep} has emerged as a powerful tool for extracting data features and has been applied in many fields, such as image processing and speech recognition. Network science researchers have applied deep learning models, including convolutional neural network~\cite{krizhevsky2012imagenet}, autoencoder neural network~\cite{hinton2006reducing}, recurrent neural network~\cite{mikolov2010recurrent}, and generative adversarial network~\cite{goodfellow2014generative}, into graph-structured data and proposed the graph neural network (GNN) models~\cite{kipf2016semi,chang2015heterogeneous,zugner2018adversarial,yu2018learning,seo2018structured}. In recent, many researchers try to introduce deep learning models, such as reinforcement learning, adversarial methods to graph learning~\cite{bojchevski2019adversarial,mendonca2019graph}. In this survey, we provide a brief introduction to traditional representation algorithms and then mainly review  NRL algorithms associated with deep learning models.

\begin{figure*}[]
  \centering
  \includegraphics[width=6.0in,height=3.4in]{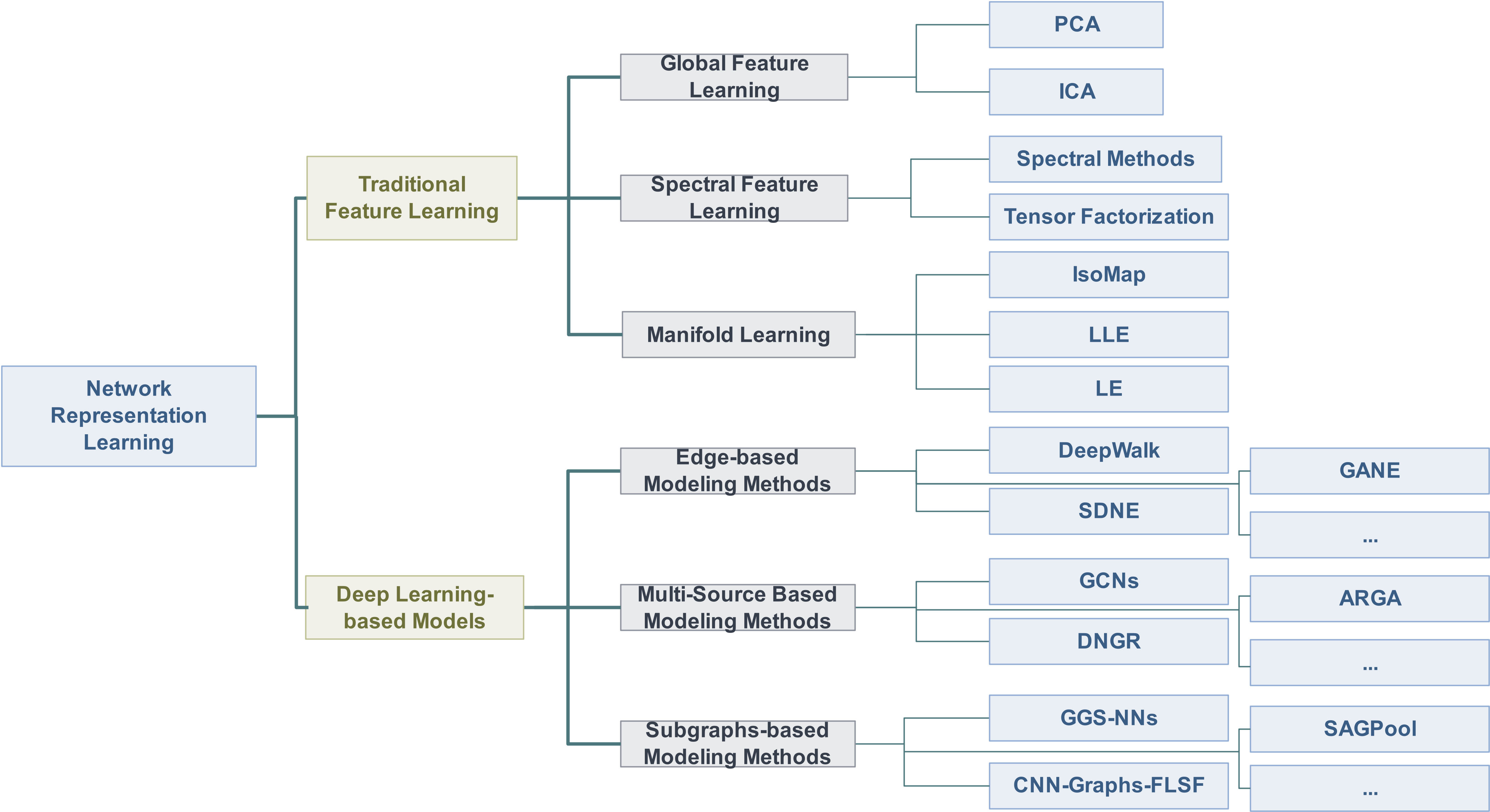}
  \caption{Categories of network representation learning algorithms.}
  \label{m_framework}
\end{figure*}

Representation learning models can be divided into ``shallow'' model and ``deep'' model. In this survey, we try to go through the development of data representation in graph-structured data from TFL to recent NRL based on deep learning. We do not intend to thoroughly summarize the various types of representation learning models in the literature. We will instead review TFL and the state-of-the-art representation learning methods mainly focusing on NRL technologies and network embedding algorithms. For understanding related algorithms, we first introduce the Word2vec framework as a basic model used by a large number of NRL algorithms to help understand what NRL is and the relationship between NRL and deep learning. Furthermore, we classify NRL into two categories, including TFL models and deep learning-based models. The overall organization of the categories of network representation learning algorithms are shown as Fig~\ref{m_framework}.

\textbf{Related Surveys}. There are already a few papers that summarized the algorithms about NRL. Our survey is different from all of them. We review several traditional feature learning algorithms and classical deep learning-based NRL models. We present deep learning-based NRL models based on the graph information they are keeping. A short introduction to the other related surveys is given as follows.

Hamilton et al.~\cite{hamilton2017representation} focus on methods and applications of NRL. They discussed NRL in two main parts: embedding nodes and embedding sub-graphs. They proposed an encoder-decoder framework to organize various NRL models. Goyal et al.~\cite{goyal2018graph} mainly reviewed applications and performances of NRL, and gave detailed performance comparisons of NRL. Zhang et al.~\cite{zhang2018network} gave a comprehensive categorization of NRL, including unsupervised methods, semi-supervised methods, and methods preserving network structure or vertex labels. The survey~\cite{cui2018survey} did not review the whole variety of NRL algorithms but focused on methods for structure-preserving and property-preserving. In recent, many researchers turn attentions to graph neural network based models, and some surveys~\cite{wu2019comprehensive,zhang2020deep,bacciu2020gentle,cao2020comprehensive} specially investigated these works in the aspects of graph learning models and aggregation methods. These models are basically generated from deep learning models to graph, such as graph attention networks, graph autoencoders, and graph reinforcement learning. There are some surveys focusing on some special cases of NRL. For example, Yang et al.~\cite{yang2020heterogeneous} reviewed heterogeneous NRL with analysis over benchmark and evaluation; Xie et al.~\cite{xie2020survey} introduced   dynamic network embedding from aspects of models.

The rest of our survey is organized as follows. We first present notations and graph related concepts in Section~\ref{section:notations}. Then,  we review the models of TFL in Section~\ref{section:trafeaturelearning} and NRL based on deep learning models in Section~\ref{section:deeplearningmodels}. In the following two sections, we discuss the application of representation learning, and list several open issues. We provide our conclusions, draw our prospects of network representation learning in future research in Section~\ref{section:conclusion}. 

\section{Notations}\label{section:notations}
This section presents TFL methods and NRL models based on deep learning methods. In the following, we first give the definitions of a Graph, Network embedding, and Laplacian matrix. Also, we list some terms and notations used in this paper in Table ~\ref{notation-label}.

\textbf{Definition 1:} \textbf{Graph}. A graph $\mathbf{G =(V,E,Y)}$ is a collection consisting of nodes (vertices or points) $\textbf{V} ={\{v_1,v_2,\ldots,v_n\}}$, edges $\textbf{E} = \{e_{i,j}\}^n_{{i,j}=1}$ and labels or side information associated to nodes $\textbf{Y}$. The edges between nodes can be directed or undirected.

\textbf{Definition 2:} \textbf{Network embedding}. Network embedding includes node embedding and edge embedding. Given a graph $\mathbf{G}$, the embedding function $\mathrm{f}:\mathbf{V}\mapsto \mathbf{U}$ maps node $v_i \in \mathbf{V}$ to embedding vector $u_i \in \mathbf{U}$, where $\mathbf{V}$ represents vectors in the original space and $\mathbf{U}$ represents vectors in the projected space. Vector $u_i$ is the newly learned node representation which often has low dimensions and preserves relevant network properties.

\textbf{Definition 3:} \textbf{Laplacian matrix}. In the graph theory, Laplacian matrix is a matrix representation of a graph. The Laplacian matrix of a simple graph is represented as: $\mathbf{L = D-A}$, where $\mathbf{D}=\mathrm{diag}(\sum_{j:j\neq i}\mathbf{W}_{ij})$ is the degree matrix, $\mathbf{W}_{ij}$ is the weight between node $i$ and $j$, and $\mathbf{A}$ is the adjacency matrix of a graph.
\renewcommand\arraystretch{1.5}
\begin{table}[!h]
\centering
\caption{Terms and Notations}
\label{notation-label}
\begin{tabular}{|p{1.5cm}<{\centering}|p{6cm}<{\centering}|}
\hline
$\textbf{G}$ & The representation of a graph includes nodes and edges \\ \hline
$\textbf{V}$  &   The set of vertices of a graph \\ \hline
$v$  &   A vertice of a graph \\ \hline
$\textbf{E}$  &   The set of edges of a graph  \\ \hline
$\textbf{L}$  &   Laplacian matrix of a graph \\ \hline
$\textbf{D}$  &   The degree matrix of a graph \\ \hline
$\textbf{A}$  &   Adjacency matrix of graph   \\ \hline
$i,j$&   The index of vertices  \\      \hline
$\textbf{u}$  &   The new representation vector   \\ \hline
$\textbf{W}$  &   The weight matrix of a graph  \\ \hline
$\mathcal{L}$  & The symbol of loss function \\ \hline
\end{tabular}
\end{table}

Representation Learning relies on an essential assumption in the manifold hypothesis~\cite{bengio2013representation}, which refers to the real-world high dimensional data, such as images with two-dimensional manifolds embedding in the high dimensional space. Graph-structured data often has high dimensions and various types. Based on the manifold hypothesis, NRL algorithm could reduce the dimensions of graph data but keeps the internal relationship of nodes.

\section{Traditional Feature Learning Models} \label{section:trafeaturelearning}
Learning the intrinsic characteristics of data is always an enormous requirement for data science. In the past decade, a great deal of traditional feature learning or representation learning algorithms have been proposed in the machine learning domain such as kernel PCA~\cite{pearson1901liii} and kernel k-means~\cite{dhillon2004kernel}. They are a set of techniques allowing a system to automatically learn the latent features from raw data. These techniques are different from feature engineering which manually sets feature parameters. In this section, we mainly focus on TFL on graphs, and we separate them into three parts: global feature learning models, spectral learning based models, and manifold learning models. Global feature learning mainly focuses on preserving global information of data. Manifold learning aims to preserve local features and information, i.e., preserving property with the neighborhood of each node in a network.
\subsection{Global Feature Learning}
As mentioned above, global feature learning methods primarily focus on preserving global information of raw data in the learning feature space. In the following, we will present several algorithms belonging to this category.

\subsubsection{PCA Algorithms}
PCA algorithm~\cite{pearson1901liii}, is one of the earliest and most popular methods used to reduce the dimensions of the data. PCA is a linear, unsupervised, generative, and global feature learning method. There are lots of global feature learning algorithms, e.g., the variants of PCA, including sparse PCA~\cite{zou2006sparse} and robust PCA~\cite{candes2011robust}. PCA can be used in NRL, such as dimension-reduced, and low-rank recovery of data. It can be applied for network visualization and clustering in network science as well.

The classical PCA algorithm has several weaknesses, e.g., it lacks the ability to scale well to a number of data samples and is sensitive to outliers. In general, the reason that the classical PCA algorithm is susceptible to outliers data is caused by the quadratic term. Robust PCA can well overcome these shortcomings of classical PCA algorithm mentioned above and it is robust to occlusions and missing values by recovering the low-rank representation~\cite{candes2011robust}. With the graph smoothness assumption, Shahid et al.~\cite{shahid2015robust} incorporated spectral graph regularization into robust PCA algorithm to improve the quality of clustering and dimension-reduced. The normalized graph Laplacian is defined as:
\begin{equation}
\mathbf{L} = \mathbf{D}^{-1/2}(\mathbf{D-A})\mathbf{D}^{-1/2} = \mathbf{I} -
\mathbf{D}^{-1/2}\mathbf{AD}^{-1/2},
\end{equation}
where $\mathbf{D} =$ diag$(d_i)$ is the degree matrix and $\mathbf{A}$ is an adjacency matrix. In general, graph-Laplacian~\cite{belkin2002laplacian} is often combined with PCA methods~\cite{feng2017joint,feng2017pca,sun2017protein} for feature extraction in the area of bioinformatics. The final experiments showed that the model can achieve better performance than the state-of-the-art models for the clustering and low-rank recovery tasks.

To accelerate the speed of robust PCA, Shahid et al.~\cite{shahid2016fast} proposed fast robust PCA (FRPCA) algorithms, which has the same advantages as the previous robust PCA. FRPCA has lower computational complexity for processing large scale datasets ($O(nlog(n))$) by using the FLANN
library~\cite{muja2014scalable}. Motivated by the emerging field of signal processing on graphs~\cite{shuman2013emerging}, FRPCA adopted the idea of graph similarity, including feature similarity and data similarity, to enhance the clustering quality in the new representation space. One problem of PCA algorithms is its high computational complexity. Here, FRPCA utilized Fast Iterative Soft Thresholding Algorithm (FISTA)~\cite{beck2009fast} to solve this problem.

There are still many variants of PCA, such as sparse PCA (SPCA)~\cite{zou2006sparse} which extends classical PCA algorithm by adding sparsity constraint on input variables and can be well applied for multivariate datasets. Megasthenis et al.~\cite{asteris2015stay} introduced a variant of SPCA which accommodates the graph constraints to analyze financial data and the data in neuroscience. Min et al.~\cite{min2018edge} proposed Edge-group Sparse PCA (ESPCA) which combines the prior gene network with the PCA method for dimension-reduced and feature extraction.

\subsubsection{ICA Algorithms}
Independent component analysis (ICA)~\cite{lee1998independent} is a classical and powerful tool in signal processing, and also has been used to analyze the structured graph data. It is widely used to brain network analysis~\cite{park2014graph,lopez2015multiple,ribeiro2017method}. Park et al.~\cite{park2014graph} proposed a variant ICA, Graph ICA, to explore the changes of cognitive networks in the brain after completing a task. The concepts of Graph-ICA can be shown as follows:
\begin{equation}
g = [g_1,\ldots,g_M]^\top = \textbf{W}[s_1,\ldots,s_M]^\top =  \textbf{W}s,
\end{equation}
where weight matrix $\textbf{W}$ represents the relationship strength between source (graph) $s$ to compose $g$. The main ideas of the algorithm are to decompose measured graphs into common source graphs and then find these canonical network components from limited sets of data in neuroimaging. Diana et al.~\cite{lopez2015multiple} directly utilized the ICA algorithm to extract features of different brain networks on the fMRI data and found the relationship among word learning with different parts of brain networks.

Ziegler et al.~\cite{ribeiro2017method} designed a method combined with the ICA algorithm and then applied it for analyzing the resting-state fMRI. In short, the authors first used ICA to deal with neuronal components and then reconstructed them as weigh graphs. The method of calculating edge weighting in the graphs depends on the contribution to the specific component.

\subsection{Spectral Learning on Graphs}
The spectral learning, which is one kind of machine learning algorithm based on spectral methods, utilizes information in the eigenvectors of the target matrix to extract hidden structure. Most of the methods based on spectral learning just consider the structural information, so they could not apply to networks with complex information. In the following, we will discuss several common spectral learning frameworks, including spectral methods, singular value decomposition, and tensor factorization.
\subsubsection{Spectral Methods}
Spectral methods are fundamental for solving problems in engineering, applied mathematics, and statistics. More specifically, network researchers have used spectral methods to solve problems in network science, such as analyzing and visualizing networks.

Community detection is a popular problem in social networks. To address this problem, Newman et al.~\cite{newman2013spectral} proposed a spectral method as they find that tradition community detection methods based on maximum modularity and likelihood methods can be treated as spectral algorithms. The main idea of the algorithm is to utilize the matrix eigenvectors to represent the networks. The proposed algorithm contains three parts: (1) modularity maximization, (2) degree-corrected block model, and (3) normalized-cut graph partitioning. Zhang et al.~\cite{zhang2014detecting} also applied the spectral method to detect communities while focused on overlapping communities in social networks. They adopted the K-medians algorithm to address the overlaps for graph clustering. In the real world, data are collected in different forms with various features and different structures. In order to deal with this situation, Li et al.~\cite{li2015large} designed a spectral clustering algorithm based on the bipartite graph. The solution is the first k smallest eigenvectors. Moreover, the authors used a fast approximation algorithm to reduce the cost of computation of multi-view spectral clustering, so as to face the requirements of large-scale graph construction. Some researchers try to combine spectral methods with deep learning~\cite{pfau2018spectral} by adopting stochastic optimization, which is successfully used to find meaningful subgoals in reinforcement learning environments.

\subsubsection{Tensor Factorization}
The knowledge graph is a hot topic in network science, which can express data or information in the form of a graph with edges representing relations and nodes representing entities. It can be applied for recommender systems, search engines, etc. A group of researchers have applied tensor factorization methods to study NRL. Trouillon et al.~\cite{trouillon2016complex} designed a method for link prediction called ComplEx, which is linear in both space and time. Besides, the algorithm exploited complex embeddings and utilized Hermitian dot product. Therefore, it is much simpler than neural tensor networks and holographic embeddings. Also, the algorithm is suitable for large datasets. Trouillon et al.~\cite{trouillon2017knowledge} extended previous work~\cite{trouillon2016complex} and utilized factorization models to study the knowledge graph. Moreover, the authors gave several proofs for the proposed model and more experiments, especially related to the training time of the models. The main idea of the algorithm is decomposing tensors into a product of embedding matrices with much lower dimensions.

\subsection{Manifold Learning}
Manifold learning methods focus on preserving local similarity among data when the new representations are learned. In recent years, this algorithm is often used to deal with network analysis tasks. We will present several manifold learning algorithms, such as Isomap, Local Linear Embedding (LLE), and Laplacian Eigenmaps (LE). They are all based on the graph construction by exploiting manifold learning.

\subsubsection{Isomap}
Similar to PCA, Isomap algorithm~\cite{tenenbaum2000global} is a classical-dimension reduced approach, which is building on classical Metric Multidimensional Scaling (MDS)~\cite{borg2003modern}. It is more powerful than other classical reduction methods as it could keep the nonlinear relations of original source. The main ideas of Isomap contain three steps: (1) constructing a neighborhood graph by using connectivity algorithm (such as KNN) from adjacency matrix; (2) computing the shortest path of entries as the geodesic distance; (3) finally, using MDS algorithm to obtain coordinate vector. The objective function is shown as follows:
\begin{equation}
\mathrm {min} \quad \Sigma_{i\neq j,\ldots,N} (d_{i,j}-||u_i-u_j||)^2,
\label{equ:isomap}
\end{equation}
where $d$ represents the shortest path obtained from step (2), and $u$ is the new representation vector that can be learned when minimizing Eq. (\ref{equ:isomap}). From Eq. (\ref{equ:isomap}), we can see that the optimized objective function is to make the distance between nodes in the new learned space similar to the distance in the original space. In other words, new low dimensional vectors approximately preserve the geodesic distance of the original data in the high dimensional space.

\subsubsection{Local Linear Embedding}
Local Linear Embedding (LLE)~\cite{roweis2000nonlinear} is an another classical nonlinear dimension-reduced approach. This algorithm relies on the manifold hypothesis, and each node lies on its neighbors. Node features can be obtained from the summation of neighbor features, so that the algorithm has the ability to preserve the locally linear structure of neighborhood. Although LLE could preserve the structural information, it could only be used to undirected graph. LLE includes three main steps: (1) selecting neighbors for each node; (2) computing the weight $\mathbf{W}_{ij}$ which is the edge weight between the node
and its neighbors:
\begin{equation}
\mathrm {min} \quad \Sigma_i ||x_i -\Sigma_j \mathbf{W}_{ij}x_j||^2,
\end{equation}
(3) computing the new low dimensional representation from weights obtained
from step (2), which is expressed as:
\begin{equation}
\mathrm {min} \quad \Sigma_i ||u_i - \Sigma_j \mathbf{W}_{ij}u_j||^2.
\label{equ:LLE}
\end{equation}

When minimizing (\ref{equ:LLE}), we can obtain the representation matrix $\textbf{U}$. In summary, LLE encodes the local information at each point
into the reconstruction weights of its neighbors and then uses these weights to compute the low dimensional embeddings.

\subsubsection{Laplacian Eigenmaps}
Laplacian Eigenmaps (LE)~\cite{belkin2002laplacian} is a popular approach to
find the low dimensional representation. Similar to the first step of LLE, LE
first constructs a graph $\mathbf{G}$ by using the k nearest neighbors, and then uses
the graph $\mathbf{G}$ to derive Laplacian matrix $\mathbf{L = D-W}$, where the
weight matrix $\textbf{W}$ is generated by heat kernel method. The authors defined an
objective function that makes connected points stay closer
to each other, which is expressed as:
\begin{equation}
\Sigma ||u_i-u_j||^2\mathbf{W}_{ij} = \mathrm{tr}(\mathbf{U}^\top\mathbf{LU}).
\end{equation}
When we minimize the above equation, the new representation matrix $\mathbf{U}$ can be obtained. In addition, TFL methods,
such as Isomap, LLE, and LE, are all just applied for the undirected graph without external node
information and focus on local features of the graph. However, they are not suitable
for large-scale networks because obtaining eigenvector from large scale matrices has
high computational complexity both in time and space.

\begin{figure*}[htbp]
  \centering
  \includegraphics[width=5.8in,height=1.6in]{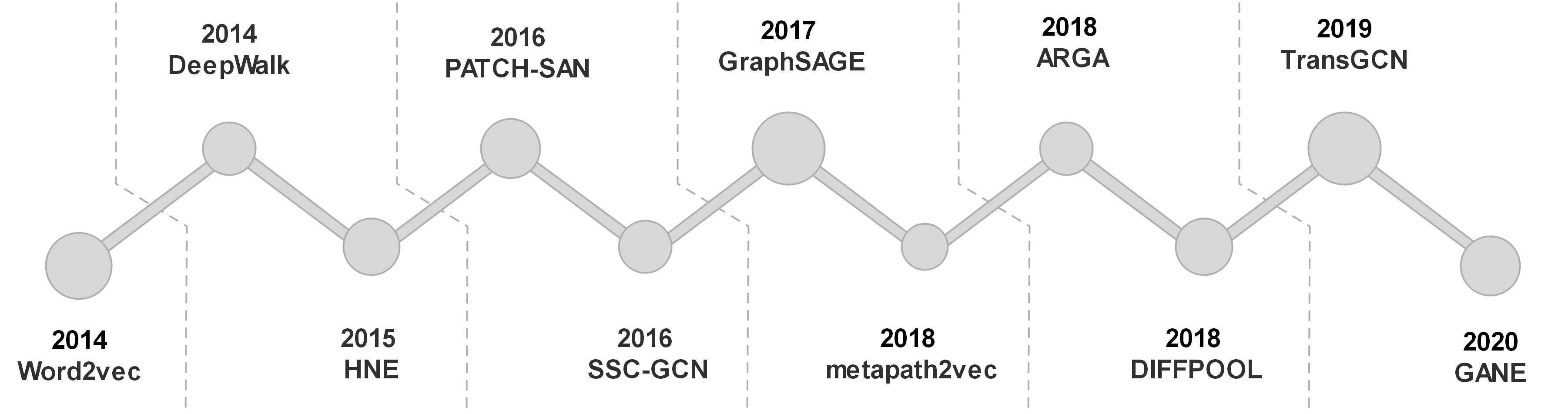}
  \caption{A timeline of some representative network representation learning methods.}
  \label{timeline}
\end{figure*}

\section{Deep Learning-based models} \label{section:deeplearningmodels}
We have witnessed the superior performance of deep learning in many fields, and they have been widely applied for image classification, speech recognition, and object detection, etc. The deep architecture can extract latent information layer by layer from data, which contributes to the performance of data processing. More precisely, the original data is transformed by a nonlinear model to a more higher-level feature representation so as to achieve more abstract representation of data. There are several deep learning-based NRL models~\cite{niepert2016learning,wang2016structural,wang2018graphgan,kumar2019predicting,liu2019shifu2} proposed in recent years. The timeline of some representative methods of them are illustrated as Fig~\ref{timeline}. Even though most of them are based on advanced models of deep learning, there have some methods having connections with traditional feature learning. We will introduce these in the related parts. In the following, we will focus on reviewing them from three subsections, as outlined in Table~\ref{algorithms_category}.

\subsection{A Taxonomy of Deep Learning-based NRL Models}
The deep learning-based NRL models have different categories. We assign them into three categories: (1) Edge-based Modeling Methods, (2) Multi-source Based Modeling Methods, (3) Subgraphs Based Modeling Methods. As deep learning-based NRL models are the major concern in this paper, we discuss and give a brief summary of them in Table~\ref{algorithms_category}. In the following subsections, We detail the characteristics of each algorithm belonging to the listed categories and provide a summary of them.

\begin{table*}[htbp]
  \centering
  \caption{A summary of NRL algorithms according to the information they are preserving}
   \resizebox{\textwidth}{!}{
    \begin{tabular}{|c|c|c|c|c|}
    \hline
    \textbf{Category} & \textbf{Algorithms } & \textbf{Neural components} & \textbf{Proximities} & \multicolumn{1}{p{15.815em}|}{\textbf{Strucutre Sequence Technology }} \\
    \hline
    \multicolumn{1}{|c|}{\multirow{9}[18]{*}{Edge-based Modeling Methods}} & Word2vec~\cite{mikolov2013efficient} & Neural Probabilistic Language Model & None & \multirow{5}[10]{*}{Random walk} \\
\cline{2-4}        & DeepWalk~\cite{perozzi2014deepwalk} & \multirow{2}[4]{*}{Skip-Gram Model} & First-order, Second-order &  \\
\cline{2-2}\cline{4-4}        & node2vec~\cite{grover2016node2vec} &     & \multirow{4}[8]{*}{Second-order, Higher-order} &  \\
\cline{2-3}        & AIDW~\cite{dai2018adversarial} & \multirow{3}[6]{*}{GANs} &     &  \\
\cline{2-2}        & GraphGAN~\cite{wang2018graphgan} &     &     &  \\
\cline{2-2}\cline{5-5}        & GANE~\cite{hong2019gane} &     &     & \multirow{2}[4]{*}{None} \\
\cline{2-4}        & LINE~\cite{tang2015line} & None & \multirow{2}[4]{*}{First-order, Second-order} &  \\
\cline{2-3}\cline{5-5}        & SDNE~\cite{wang2016structural} & \multirow{2}[4]{*}{Autoencoder} &     & Adjacency matricx \\
\cline{2-2}\cline{4-5}        & DNGR~\cite{cao2016deep} &     & Second-order, Higher-order & Surfing model \\
    \hline
    \multicolumn{1}{|c|}{\multirow{6}[12]{*}{Multi-source Based Modeling Methods}} & HNE~\cite{chang2015heterogeneous} & \multirow{3}[6]{*}{(Graph)Convolutional Neural Network} & \multirow{2}[4]{*}{None} & Linear transformation matrices \\
\cline{2-2}\cline{5-5}        & PATCH-SAN~\cite{niepert2016learning} &     &     & Graph normaliztion \\
\cline{2-2}\cline{4-5}        & SSC-GCN~\cite{kipf2016semi} &     & \multicolumn{1}{l|}{First-order} & \multicolumn{1}{l|}{Layer-wise propagation rule} \\
\cline{2-5}        & Planetoid~\cite{yang2016revisiting} & Feed-forward neural networks & Second-order, Higher-order & \multirow{2}[4]{*}{None} \\
\cline{2-4}        & TransNet~\cite{tu2017transnet} & Autoencoder & \multirow{4}[8]{*}{None} &  \\
\cline{2-3}\cline{5-5}        & ARGA~\cite{pan2018adversarially} & Autoencoder, GANs &     &  Adjacency matrix  \\
\cline{1-3}\cline{5-5}    \multicolumn{1}{|c|}{\multirow{5}[10]{*}{Subgraphs Based Modeling Methods}} & GGS-NNs~\cite{li2015gated}  & Graph Neural Networks &     & \multirow{2}[4]{*}{None} \\
\cline{2-3}        & CNN-Graphs-FLSF~\cite{defferrard2016convolutional} & Convolutional Neural Network &     &  \\
\cline{2-5}        & DIFFPOOL~\cite{ying2018hierarchical} & \multirow{3}[6]{*}{Graph Neural Network,Pooling} & \multirow{3}[6]{*}{First-order} & \multirow{3}[6]{*}{Layer-wise propagation rule} \\
\cline{2-2}        & HGP-SL~\cite{zhang2019hierarchical} &     &     &  \\
\cline{2-2}        & SAGPool~\cite{lee2019self} &     &     &  \\
    \hline
    \end{tabular}%
    }
  \label{algorithms_category}%
\end{table*}%

\subsection{Edge-based Modeling Methods}
Graph-structured data is a complex data type containing edges and nodes. In
real world, graph edge can represent the link between users and products or links
of friends. Lots of NRL algorithms just consider the structure of the graph, such as
first-order proximity of nodes and second-order proximity of nodes. We cluster
these NRL models by focusing on the graph structure as edge-based modeling
methods. In addition, there are several NRL models based on Skip-Gram
model~\cite{mikolov2013efficient}, which is a powerful model in natural language
processing. Moreover, the random walk approach~\cite{xia2019random} has been applied to capture graph structure.
To understand these NRL algorithms deduced from Skip-Gram model, we
start with a brief introduction of Word2vec model~\cite{mikolov2013efficient} in this subsection.

\subsubsection{Word2vec}
Word2vec model~\cite{mikolov2013efficient}~\cite{morin2005hierarchical} is
recognized as a powerful tool in natural language processing. It can
reconstruct one-hot vector representations of words (word embedding).
Actually, the framework of the model is a variant of neural probabilistic
language model with three layers, and the word embedding representations are the
matrices between input layer and hidden layer. After that, several
literatures~\cite{goldberg2014word2vec,levy2014neural,rong2014word2vec} have
been proposed other variants to explain the principle of Word2vec. Levy et
al.~\cite{levy2014neural} pointed out that the neural word embedding is one kind
of implicit matrix factorization. Given training words set  \{$w_1$,
$w_2$, $w_t$, \ldots, $w_n$\}, where $t$ is the position in a text and the aim
of Word2vec is to learn an estimated model:
\begin{equation}
\mathrm{F}(\Theta,w_t,w_{t-1},\ldots,w_{t-n+1})=\mathrm{P}(w_t|w_{t-1},\ldots,w_{t-n+1}),
\end{equation}
where function $\mathrm{P}$ is the conditional probability, function $\mathrm{F}$ is the function carried out
by using a neural network and its free parameters, and $\Theta$ denotes the feature
vector matrix (neural network matrix). The weight matrices can
be learned by training the model when maximizing the empirical conditional
probability of model $\mathrm{F}$:
\begin{equation}
\mathrm {max} \quad\sum_t \mathrm{F}(\Theta,w_t,w_{t-1},\ldots,w_{t-n+1}).
\end{equation}

\begin{figure}[!htb]
  \centering
  \includegraphics[width=3in,height=1.8 in]{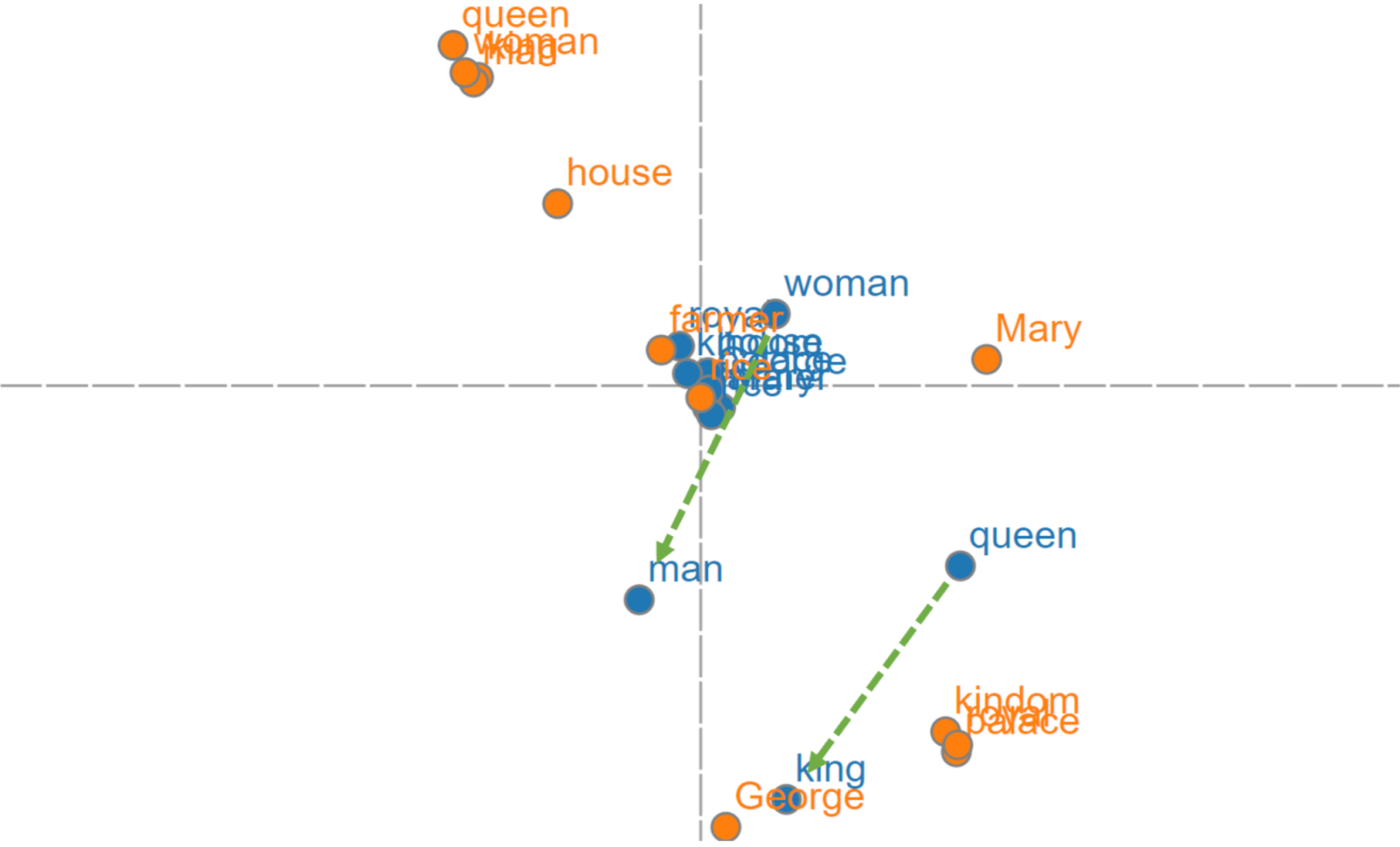}
  \caption{An example of word vectors embedding in two dimensional space.}
  \label{king_queen}
\end{figure}

There are some variants of the model based on the way of normalization, e.g.,
softmax normalization and hierarchical softmax. The word vector is attracting
interest due to the feature that semantic similarity words are located close to
each other in the word vector space (representation space). In addition, the
vectors can be computed by linear mathematical operations, for example,
``king''-``queen''=``man''-``women'' as shown in Fig.~\ref{king_queen}.
These intriguing property of the word vector shows that it contains semantic
information existing in the real world.

\subsubsection{DeepWalk}
DeepWalk~\cite{perozzi2014deepwalk} is a NRL algorithm, and it can learn latent representation of vertices in networks. The algorithm is the first generalization of Word2vec to networks. The truncated random walk approach is utilized to capture the graph structure, and then generate a sequence of vertices. However, the random walk approach is unbiased, which means it can not conduct breadth-first search or depth-first search on graph based on preferences. That provides a chance to improve the embedding performance by node2vec~\cite{grover2016node2vec}. The frequency of the vertex appears in the sequences following power-law distributions, which is similar to the distributions of words in natural language. This is the main reason that Word2vec algorithm can be used to generalize the network-structured data. Given a random walk sequence $v_1,v_2,\ldots, v_l$, $l$ is the length of word sequence. The training objective of DeepWalk is the same to the Word2vec algorithm. Given a previous vertex $v_i$, the likelihood of observing vertices $v_{i-w},\ldots, v_{i-1}, v_{i+1},\ldots, v_{i+w}$ is expressed as
\begin{equation}
\mathrm{P}(v_{i-w},\ldots,v_{i-1},v_{i+1},\ldots,v_{i+w}|v_i).
\end{equation}

Now, learning an effective vertex representation $\Phi(v_i)$ ($\Phi(v_i) =\Theta^T\cdot v_i$) becomes an optimization problem:
\begin{equation}
\mathrm{max} \quad \mathrm {log}\ \mathrm{P}({v_{i-w},\ldots,v_{i-1},v_{i+1},\ldots,v_{i+w}}|\Phi(v_i)).
\end{equation}

Different from the form of the adjacency matrix, vertex vector representation can avoid the data sparse problem, which can achieve higher computational efficiency. Furthermore, the random walk approach is leveraged to generate sequences of vertices based on local information. This characteristic enables DeepWalk to run on the distributed systems so as to meet the requirement of large-scale data processing. 

\subsubsection{Node2vec}
\begin{figure}[]
	\centering
	\includegraphics[width=2.1in,height=1.9in]{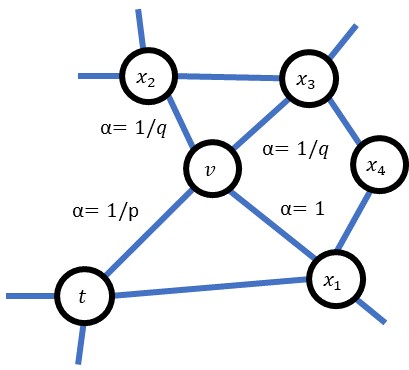}
	\caption{Node2vec random walk strategy.}
	\label{node2vec}
\end{figure}

Analogous to DeepWalk based on Word2vec, node2vec~\cite{grover2016node2vec} algorithm was proposed by extending the Skip-gram architecture~\cite{mikolov2013efficient} to networks. The algorithm introduced a flexible neighborhood sampling strategy than DeepWalk, which captures network structure controlled by two hyperparameters $p$ and $q$. They are used to interpolate random walk with breadth-first sampling or depth-first sampling. Given several nodes $t,v,x_1,x_2,x_3$ as shown in Fig.~\ref{node2vec}, the unnormalized transition probability between $v$ and $x$ is decided by $\alpha_{\mathrm{pq}}(t,x)\cdot w_{\mathrm{vx}}$. The piecewise function $\alpha_{\mathrm{pq}}(t,x)$ is expressed as
\begin{equation}
\alpha_{\mathrm{pq}}(t,x)=\left\{
\begin{aligned}
\frac{1}{p}\quad \mathrm{if}\ d_{\mathrm{tx}} = 0\\
1          \quad \mathrm{if}\ d_{\mathrm{tx}} = 1\\
\frac{1}{q}\quad \mathrm{if}\ d_{\mathrm{tx}} = 2,
\end{aligned}
\right.
\end{equation}
where $w_{\mathrm{vx}}$ is the static edge weight and $d_{\mathrm{tx}}$ denotes the shortest path distance between nodes $t$ and $x$. Actually, the unbiased random walk strategy of DeepWalk is a special case of node2vec with $p = 1$ and $q = 1$. When tuning these parameters, the model can trade off the preference of focusing on the local structure or the global structure and therefore learns high quality and more information embeddings compared with DeepWalk.

\subsubsection{LINE}
Large-scale Information Network Embedding (LINE) algorithm~\cite{tang2015line} is not a deep learning based model, but is often compared with DeepWalk and node2vec~\cite{grover2016node2vec}. Qiu et al.~\cite{qiu2018network} pointed out that LINE, Node2vec and LINE can be implicitly categorized as matrix factorization frameworks. LINE is able to preserve the first-order proximity and second-order proximity. But it could not preserve the high-order proximity like node2vec. The first-order proximity refers to the proximity of two nodes connected with one-hop, and it can be measured by the joint probability distribution:
\begin{equation}
\mathrm{p}_1(v_i,v_j)=\frac{1}{1+\mathrm {exp}(-\vec{u_i}^\top \cdot \vec{u_j})},
\end{equation}
where $\vec{u_i}$ and $\vec{u_j}$ stand for the vector representation of the nodes $v_i$ and $v_j$, respectively. The second-order is similar to the first-order, but considers two nodes with a range of two-hop. Its proximity is the probability of the context node $v_j$ generated by node $v_i$, i.e.,
\begin{equation}
\mathrm{p}_2(v_j|v_i)=
\frac{\mathrm {exp}(\vec{u_j}'^\top \cdot\vec{u_i})}
{\sum_k \mathrm {exp}(\vec{u_k}'^\top \cdot
\vec{u_i})}.
\end{equation}
The second-order proximity means that nodes with similar distribution are similar to each other. To preserve the first-order proximity or the second-order proximity, the optimization objective of the algorithm tries to minimize the loss functions derived from KL-divergence between probability distribution and empirical distribution.

\subsubsection{SDNE}
\begin{figure}[]
  \centering
  \includegraphics[width=3.1in,height=1.7in]{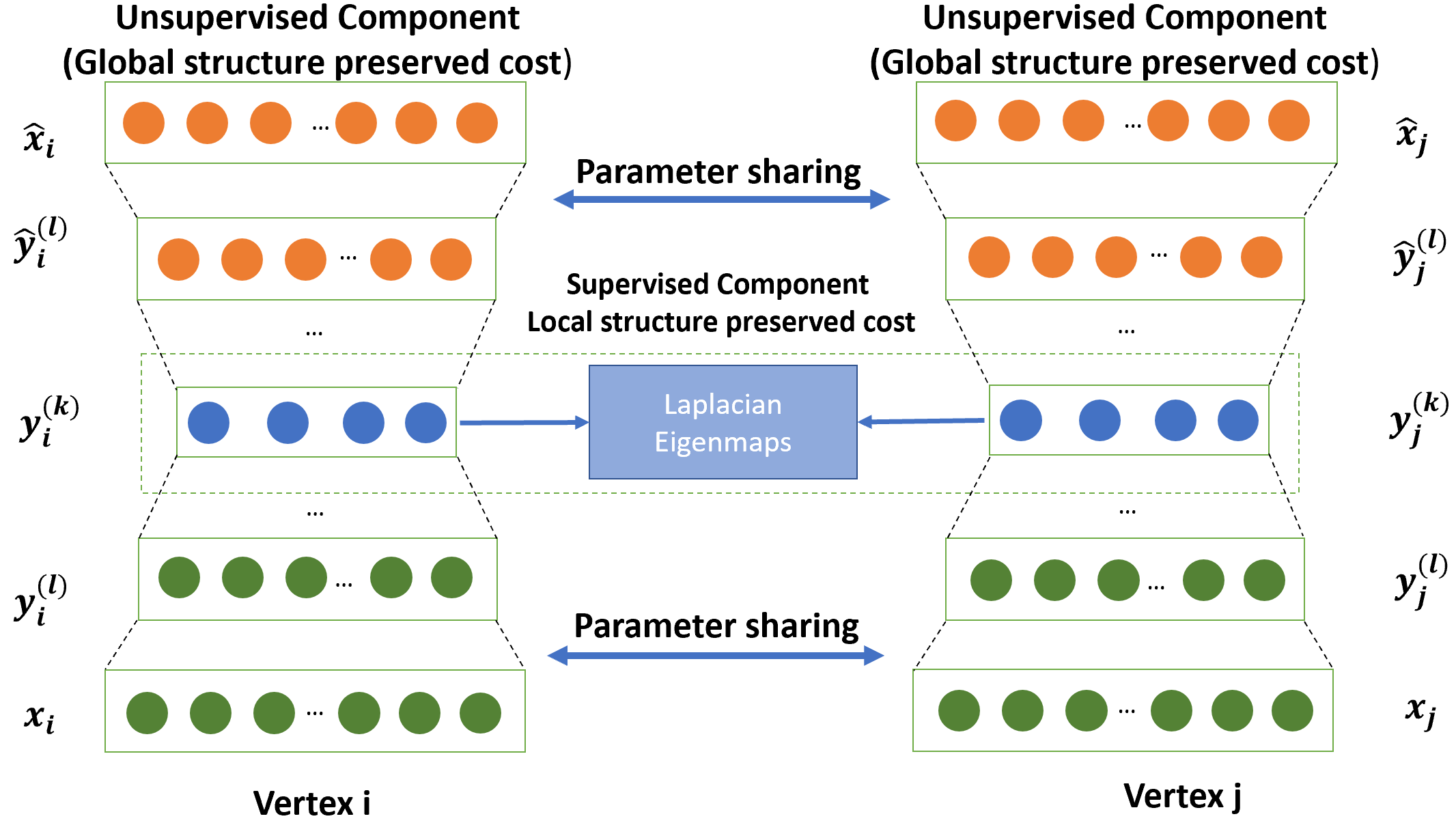}
  \caption{The structure of deep network embedding.}
  \label{sdne}
\end{figure}

Most NRL algorithms cannot extract the high nonlinear network-structured feature. Wang et al.~\cite{wang2016structural} designed a semi-supervised model named Structure Deep Network Embedding (SDNE), which is a representative NRL model based on deep autoencoder approach~\cite{salakhutdinov2009semantic}. The framework of SDNE is shown in Fig.~\ref{sdne}. Similar to LINE which focuses on the graph-structured proximity of nodes, SDNE also preserves the first-order and second-order proximity of nodes. To address these structure-preserving and sparsity problems, the basic ideas of the algorithm are stated as two parts: (1) utilizing unsupervised component combining with deep autoencoder to preserve second-order proximity, which means that vertices with similar neighborhood stay close in the latent representation space; (2) using first-order proximity as the supervised information to make similar vertices more similar in the embedding space, where the objective function is based on Laplacian eigenmaps~\cite{belkin2003laplacian}. The loss function for second-order proximity is given by:
\begin{equation}
\mathcal{L}_{\mathrm{2nd}} = \sum^{n}_{i=1}||(\hat{r}_i-r_i)\odot b_i||_2^2,
\end{equation}
where $\hat{r}_i$ denotes the reconstructed representation, and $r_i$ is the input representation representing the neighborhood structure of the vertex. Notation $\odot$ represents the Hadamard product and $\mathbf{b_i} = \{b_{i,j}\}_{j=1}^n$ is used to impose more penalty to the reconstruction error of non-zero elements than zero elements, where
\begin{equation}
\mathrm{b}_{i,j}=
\begin{cases}
\beta>1 & \text{$s_{i,j}=0$},\\
1&        \text{otherwise}.
\end{cases}
\end{equation}
SDNE can preserve the local structure of network. The first-order proximity is adopted to represent the local network structure and the loss function is expressed as
\begin{equation}
\mathcal{L}_{\mathrm{1st}} = \sum^{n}_{i,j=1} s_{ij}||y_{i}^{(K)}-y_{j}^{(K)}||^2_2,
\end{equation}
where $s_{ij}$ is an instance from the the adjacency matrix $\mathbf{S}$, and $y_i$ is the latent representation of node. As mentioned above, the objective of the above loss function is to make similar vertices more similar in the embedding space by utilizing the supervised information. The two loss functions are all distance-based model similar to Isomap~\cite{tenenbaum2000global}, LE~\cite{belkin2002laplacian}, and LLE~\cite{roweis2000nonlinear}. To preserve the first-order and second-order proximity, the mix loss function combined
$\mathcal{L}_{\mathrm{1st}}$ with $\mathcal{L}_{\mathrm{2nd}}$ is defined as
\begin{equation}
\mathcal{L} = \mathcal{L}_{\mathrm{2nd}}+\alpha \mathcal{L}_{\mathrm{1st}} + \nu
\mathcal{L}_{\mathrm{reg}},
\end{equation}
where $\mathcal{L}_{\mathrm{reg}}$ is a regularizer term to avoid overfitting. The final latent representation of vertices can be achieved when the above mix loss function is minimized.

\subsubsection{DNGR}
\begin{figure*}[]
  \centering
  \includegraphics[width=6in,height=2.1in]{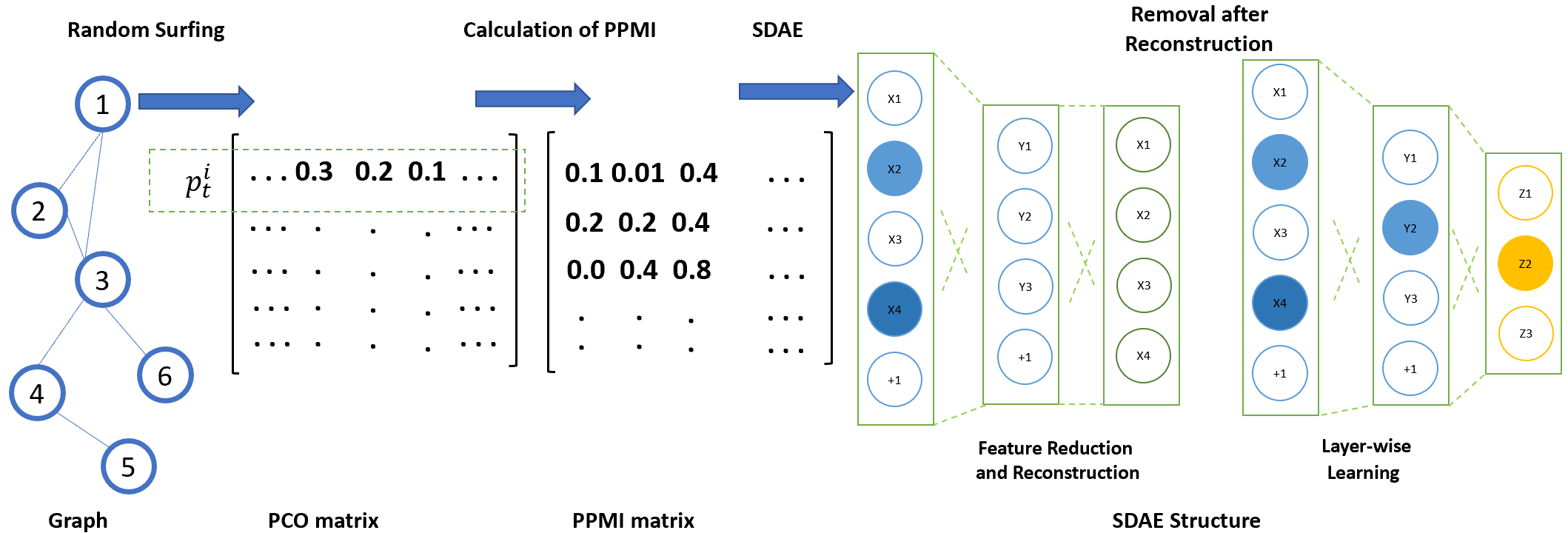}
  \caption{Deep Neural Graph Learning components.}
  \label{dngr_components}
\end{figure*}

Analogous to SDNE depending on the deep neural network model, Deep Neural Graph Learning (DNGR)~\cite{cao2016deep} is another NRL algorithm incorporating deep autoencoders with network features. In contrast to algorithms using a truncated random walk, such as DeepWalk and node2vec, the DNGR algorithm utilizes random surfing model to overcome the drawback that they cannot capture weighted graphs and cope with evolved graphs. Two important contributions are stated in~\cite{cao2016deep}: (1) designing a random surfing model motivated by PageRank mode, which can be directly applied for weighted graphs and product the probabilistic co-occurrence (PCO) matrix; (2) demonstrating a novel model for accurately learning vertex representation of weighted graphs. The main ideas of DNGR model are to transform the PCO matrix captured by random surfing model to positive pointwise mutual information (PPMI) matrix and then feed them into stacked denoising autoencoder~\cite{vincent2010stacked} so as to learn the vertex latent representation. The main processes are shown in Fig.~\ref{dngr_components}.

\subsubsection{AIDW}
Most NRL algorithms ignore the robustness of representation. To overcome this weakness, The authors~\cite{dai2018adversarial} proposed an Adversarial Inductive DeepWalk (AIDW) model consisting of structure preserving component and an adversarial learning component. AIDW could well preserve structure information while having robustness to representation. The trick behind AIDW is introducing the adversarial learning model including a discriminator and a generator (structure preserving component) to enhance the representation from the structure preserving component of AIDW. However, the adversarial learning model always has high computation time because it conducts game playing between discriminator and generator. Here, the discriminator is trained to differentiate between feature vectors and prior samples. The loss function of discriminator is expressed as:
\begin{equation}
\begin{aligned}
\mathrm{O}_D(\theta_2) = \mathbb{E}_{z\sim p(z)}[\mathrm{log}\,\mathbf{D}(\mathbf{Z};\theta_2)]+\\ \mathbb{E}_x[\mathrm{log} (1-\mathbf{D}(\mathbf{G}(x;\theta_1);\theta_2))],
\end{aligned}
\end{equation}
where $\mathbf{G}(*;*)$ is a generator. Here, the two models improve their performance by using the minimax game mechanism. 

The authors~\cite{yu2018learning} adopted a similar policy with AIDW that use GAN to enhance the performance of embedding, and proposed a novel NRL algorithm with adversarially regularized autoencoders (NetRA). Specifically, NetRA involves LSTM to product positive samples to feed generative model. However, the regularization of NetRA is static. To further improve the ability of adversarial training on graph, the authors\cite{feng2019graph} developed a framework, which could dynamically regularize with graph structure. 

\subsection{Multi-Source Based Modeling Methods}
Besides graph-structured data, there are other types of information, including vertex attributes and vertex labels, etc. We call this information as multi-source of nodes. It is no doubt that efficiently using these data can enormously improve the performance of network representation. In recent years, many studies have focused on the multi-source embedding of graphs, such as considering labels information~\cite{yang2015network} and heterogeneous network embedding~\cite{chang2015heterogeneous}. In this section, we will illustrate the NRL algorithms, which consider both graph structure and the features of nodes.

\subsubsection{HNE}
Research in NRL focuses on homogeneous networks rather than heterogeneous networks. Chang et al.~\cite{chang2015heterogeneous} designed Heterogeneous Network Embedding (HNE) algorithm to leverage deep learning architectures. Besides, HNE has several key advantages than traditional linear embedding models, such as being able to handle the dynamic networks, being suitable for network-oriented data mining applications. To transform different types of data into a uniform representation space, a relatively linear transform matrix is introduced:
\begin{equation}
\tilde{x} = \mathbf{U}^\top x,\ and \ \tilde{z} = \mathbf{V}^\top z,
\end{equation}
where $\mathbf{U}$ and $\mathbf{V}$ denote the linear transformation matrices, and $\tilde{x}$ and $\tilde{z}$ are the transformed samples. The way of linear transformation is often used to transform embedding space by traditional feature learning based on linear projection like PCA~\cite{pearson1901liii}, LLE~\cite{roweis2000nonlinear}. Importantly, to represent the similarity between two data points, the inner product is used in the projected space. Based on that, to denote relationship of heterogeneous linkages in networks, a decision function is designed
\begin{equation}
\mathrm{d}(x_i,x_j) = \mathrm{s}(x_i,x_j) -t_{II},
\end{equation}
where $\mathrm{s}(*,*)$ denotes the inner product of two samples of $x$ and $z$ respectively., and $t_{II}$ is a bias-based value. Generally speaking, most representation learning algorithms can be seen as mathematical optimization problems, and the loss function of HNE is defined as:
\begin{equation}
\mathcal{L}(x_i,x_j) = \mathrm {log}(1+ \mathrm {exp}(-\mathbf{A}_{i,j}\cdot \mathrm{d}(x_i,x_j))).
\end{equation}

Actually, the above equation can be regarded as a binary logistic regression. Another fundamental characteristic of HNE is a deep structure including a CNN structure with fully connected layers to learn features of image and text, and thus it can model complex networks with heterogeneous components. 

There are various deep learning-based methods for dealing with heterogeneous networks. For example, metapath2vec~\cite{dong2017metapath2vec} utilizes random walks method to capture graph structure information and then feeds them to HeterogeneousSkipGram to embed vectors. HAN~\cite{Wang2019Heterogeneous} involves attention mechanism to improve embedding performance. Zhang et al.~\cite{zhang2019heterogeneous} proposed a GNN-based heterogeneous networks embedding algorithm, namely HetGNN. The authors consider that GNN could capture the rich neighborhood information. Existing approaches focus primarily on static networks, while a HIN in reality is usually changing with time. Zhang et al.~\cite{wang2020dynamic} developed a dynamic heterogeneous network embedding algorithm utilizing hierarchical attentions mechanism. There are some traditional methods for heterogeneous network embedding, such as TransN~\cite{li2020transn} based on dual-learning mechanism, RHINE~\cite{shi2020rhine} based on euclidean distance and translation-based distance.

\subsubsection{Planetoid}
In the real world, most datasets are composed of unlabeled data. How to leverage a large amount of unlabeled data to improve data analysis performance is still a considerable challenge. To represent unlabeled data in the graph, Yang et al.~\cite{yang2016revisiting} designed a novel semi-supervised learning algorithm for graph embedding (Planetoid). The authors specially developed two variants methods containing transductive graph embedding and inductive graph embedding. The transductive graph embedding is applied for predicting class label and graph context based on the input feature of observed labeled data and embeddings extracted from graph structure. The loss function of transductive graph embedding is expressed as:
\begin{equation}
-\frac{1}{L}\Sigma ^L_{i = 1} \ \mathrm {log}\ \mathrm{p}(y_i|x_i,e_i)-\lambda
\mathbb{E}_{(i,c,\gamma)} \mathrm {log}\ \mathrm{\sigma}(\gamma
w_c^\top e_i),
\label{equ:transductive}
\end{equation}
where the first term is the probability of predicting labels, and the second term is the loss function for predicting graph context. To generalize unobserved instances, the inductive learning relies on the input feature $x$ and the embedding works as a parameterized function of the feature of $x$. Similar to the loss function of transductive learning, the loss function here is defined as:
\begin{equation}
-\frac{1}{L}\Sigma ^L_{i = 1} \ \mathrm {log}\ \mathrm{p}(y_i|x_i)-\lambda
\mathbb{E}_{(i,c,\gamma)} \mathrm {log}\ \sigma(\gamma w_c^\top \mathrm{h}^{l_1}(x_i)).
\label{equ:inductive}
\end{equation}

Compared with Eq. (\ref{equ:transductive}), Eq. (\ref{equ:inductive}) replaces embedding of instance $e_i$ with embedding $\mathrm{h}^{l_1}(x_i)$. In addition, Planetoid framework is based on feed-forward neural networks, of which the stochastic gradient descent (SGD) is adopted to train the model in mini-batch mode.

\subsubsection{PATHCHY-SAN}
\begin{figure}[]
  \centering
  \includegraphics[width=3in,height=1.5in]{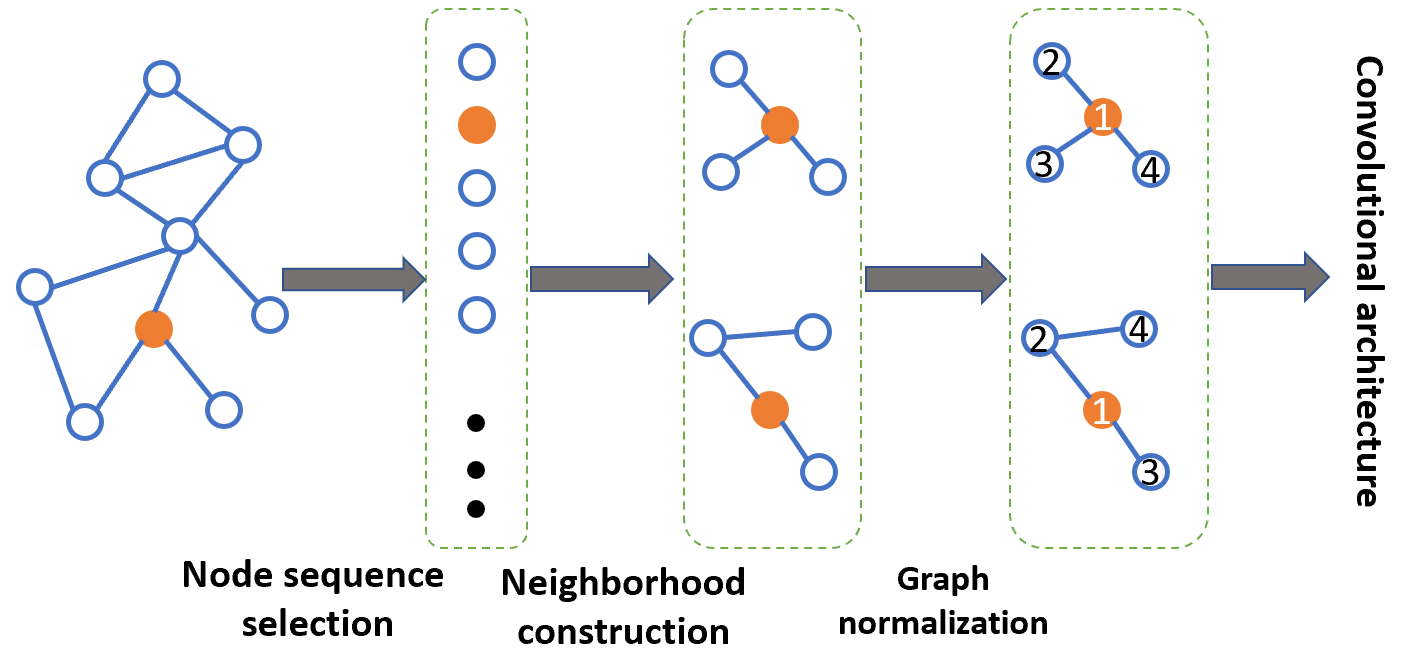}
  \caption{CNN for graph architecture.}
  \label{nn_for_graph}
\end{figure}

From the view that arbitrary graph can be seen as an image, Niepert et al.~\cite{niepert2016learning} proposed a deep learning-based algorithm: PATHCHY-SAN, for learning arbitrary graph by integrating CNN. The algorithm opens up a novel perspective that deep learning methods can be used to solve graph embedding problems. The main idea of the algorithm is transforming graph data to a special form combined with existing convolutional network components. PATCHY-SAN model contains four steps:
(1) node sequence selection;
(2) neighborhood graph construction;
(3) normalizing the extracting neighborhood graph;
(4) combining with existing CNN components,
which is illustrated in Fig.~\ref{nn_for_graph}. In step (3), in order to optimize graph normalization,  an optimal normalization problem is defined to find the optimal labeling approach $\hat{l}$, which can assign similar structural nodes to the same relative position in the adjacency matrices
for a given collection of graphs
\begin{equation}
\hat{l} = \argmin_l\
\mathbb{E}_g[|\mathrm{d}_{A}(\mathbf{A}^l(\mathbf{G}),\mathbf{A}^l(\mathbf{G}'))-\mathrm{d}_{G}(\mathbf{G,G'})|],
\end{equation}
where $g$ is a collection of unlabeled graphs, $l$ denotes an injective graph
labeling procedure, $\mathbf{A}^l(\mathbf{G})$ is a unique adjacency matrix of
graph $\mathbf{G}$ determined by labeling procedure $l$, $d_G$ denotes
the distance between graphs based on nodes and $d_A$ based on matrices. From the equation, the
optimal labeling produce $\hat{l}$ can be obtained when the expected difference
between the above two types of distance is minimized.

\subsubsection{GCNs}
Different from the above frameworks based on Word2vec, Kipf and Welling~\cite{kipf2016semi} proposed a semi-supervised graph convolutional networks (GCNs) considering  graph structure and node label information. The authors refined and optimized the previous GCN model proposed by Bruna et al.~\cite{bruna2013spectral} and successfully draw attentions of researchers to graph neural network. This previous GCNs model is based on spectral graph regularization, which is often used by traditional feature learning methods like robust PCA~\cite{shahid2015robust}. GCNs is a scalable approach, which can be directly applied for graph-structured data. The key innovation of the algorithm is introducing an effective neural network layer-wise propagation rule for graphs
\begin{equation}
\mathbf{H}^{(l+1)} = \sigma
(\tilde{\mathbf{D}}^{-\frac{1}{2}}\tilde{\mathbf{A}}\tilde{\mathbf{D}}^{-\frac{1}{2}}\mathbf{H}^{(l)}\mathbf{W}^{(l)}),
\end{equation}
where $\sigma(\cdot)$ denotes the nonlinear activation function, such as function $\mathrm{tanh}$, $\mathbf{W}^{(l)}$ is the free weight matrix of a layer, and $\tilde{\mathbf{A}}$ represents the adjacency matrix of an undirected graph with added self-connections. $\mathbf{H}^{(l)}$ is the matrix of activations in the $lth$ layer, for example, $\mathbf{H}^0 = \mathbf{X}$ (feature matrix) and $\mathbf{H}^l = \mathbf{Z}$ (The final desired feature matrix). GCNs is a differentiable generalization of the Weisfeiler-Lehman algorithm~\cite{douglas2011weisfeiler} as its propagation rule can be interpreted as a variant of a hash function of that. To achieve semi-supervised learning, graph-based regularization~\cite{zhu2003semi} is adopted to learn graph feature, and it includes two steps: (1) getting embedding of nodes; (2) training classifier on the embeddings. 

There are some GCNs-based variant methods for improving GCNs capability from different aspects. Li et al.~\cite{li2018deeper} proved that GCNs is actually a special form of Laplacian smoothing and then proposed the co-training and self-training approaches to improve the learning efficiency of GCNs framework. Chen et al.~\cite{chen2017stochastic} developed control variate based algorithms to overcome the receptive field size growing problem of GCNs so as to arrive comparable convergence speed. Hamilton et al.~\cite{hamilton2017inductive} designed an improved GCNs model: GraphSAGE with learnable aggregation functions rather than using graph Laplacian, which allows GCNs to apply to a large graph. Chen et al.~\cite{chen2018fastgcn} further improves the sampling algorithm based on GraphSAGE and obtain a better computational efficiency to a large graph. There are many other variant GCNs-based models future enhancing performances like SGCN~\cite{derr2018signed}, mGCN~\cite{ma2019multi}, and Deep-GCNs~\cite{li2019can}.

Inspired by the success of GCNs, several researchers involved this model to knowledge graph representation. Schlichtkrull et al.~\cite{schlichtkrull2018modeling} first utilized the GCN and proposed the R-GCN. The authors designed a matrix transform method to represent the relations between facts. This method could address the embedding problem caused by too many types of relationships. Cai et al.~\cite{cai2019transgcn} combined the TransE~\cite{bordes2013translating} with GCN and proposed TransGCN, which could be directly used to link prediction of the heterogeneous relations knowledge graph, while has less parameters than R-GCN~\cite{schlichtkrull2018modeling}. Wang et al.~\cite{wang2019logic} further improved the propagation model, proposed the logical attention network (LAN). The model considers the disorder and inequality nature of entities, so as to well learn relations between the entities and the corresponding neighbors.

\subsubsection{TransNet}
Realizing that there are rich semantic information on edges, Tu et al.~\cite{tu2017transnet} proposed TransNet-based NRL model to extract social relationships from networks, and the interactions between nodes can be regarded as a translation operation. Instead of utilizing CNN, the algorithm designed an auto-encoder framework to learn edge latent representation. Autoencoder is utilized to reconstruct edge labels and vertex vectors. These vertex vectors of edges stay in a continued space and we have
\begin{equation}
\mathrm{\mathbf{u}}+\mathrm{\mathbf{l}}\approx
\mathrm{\mathbf{v'}},\label{equ:TransNet}
\end{equation}
where $\mathrm{\mathbf{u}}$ and $\mathrm{\mathbf{v'}}$ denote the representations of vertices, and $\mathrm{l}$ is the edge representation derived from label set $l$. To minimize the distance $\mathbf{d}(*,*)$ at the left and right side of (\ref{equ:TransNet}), a hinge-loss is defined as
\begin{equation}
\mathcal{L}_{\mathrm{trans}}  = \mathrm{max}(\gamma
+\mathrm{d}(u+\mathrm{l},\mathrm{v'})-\mathrm{d}(\mathrm{\hat{u}}+\hat{\mathrm{l}},\mathrm{\hat{v}')},0),
\end{equation}
where $\mathbf{L}_1 $ norm is adopted, $\gamma > 0 $ is a margin hyper-parameter, and $(\hat{u},\hat{v},\hat{l})$ denotes a negative sample of original variables from the negative sampling set. Similar to SDNE framework, the deep model used by TransNet for learning edge representation is deep autoencoder, and the reconstructed loss function is a distance-based model similar to LE~\cite{belkin2002laplacian}, LLE~\cite{roweis2000nonlinear}. The loss function is expressed as
\begin{equation}
\mathcal{L}_{\mathrm{rec}} = ||(s-\hat{s})\odot \mathrm{x}||,
\end{equation}
where $s$ and $\hat{s}$ denote input and output, respectively. Finally, a joint optimization objective is defined by integrating the loss functions mentioned above
\begin{equation}
\mathcal{L} = \mathcal{L}_{\mathrm{trans}} + \alpha [\mathcal{L}_{\mathrm{ae}}(l)+
\mathcal{L}_{\mathrm{ae}}(\hat{l})]+ \eta \mathcal{L}_{\mathrm{reg}},
\end{equation}
where $\lambda$ and $\alpha$ are two hyper-parameters to regularize the
importance of different parts of the model.
\subsubsection{ARGA}
Previous works have proved that GCNs is a powerful tool to represent graph-structured data. Pan et al.~\cite{pan2018adversarially} proposed a novel adversarial graph embedding framework, namely ARGA leveraging GCNs as a graph encoder. Similar to AIDW, ARGA utilized GAN to enhance robustness of embedding while preserving structure and node label information. To keep both structure  and node label information, the authors developed a variant encoder based on GCNs, defined as follows:
\begin{equation}
\begin{aligned}
\mathbf{Z}^1 = \mathrm{f}_{\mathrm{Relu}}(\mathbf{X},\mathbf{A}|\mathbf{W}^{(0)}), \\
\mathbf{Z}^2 = \mathrm{f}_{\mathrm{linear}}(\mathbf{Z},\mathbf{A}|\mathbf{W}^{(1)}),
\end{aligned}
\end{equation}
where $\mathbf{X}$ represents the node content, $\mathbf{A}$ is the graph-structured information, such as adjacency matrix. 

There are some other GAN-based network embedding algorithms. For example, HeGAN~\cite{hu2019adversarial} utilize a generator to produce negative samples so as to achieve better embedding performance for heterogeneous information networks; Graphite~\cite{grover2019graphite} is variational autoencoders based embedding algorithm while involving iterative message passing procedure to raise performance, etc.  

\subsection{Subgraphs-Based Modeling Methods}
The NRL algorithms mentioned above just seek to address node embedding. In addition, there are some requirements for learning the representation of subgraphs or the whole graph, which refers to a set of nodes and edges, such as protein and molecules. Subgraphs embedding can be applied for learning molecular fingerprints~\cite{duvenaud2015convolutional} and predicting multicellular function~\cite{zitnik2017predicting}, etc. In addition, the fixed-size subgraphs can be treated as motifs~\cite{milo2002network,wang2020Motif} or graph kernel~\cite{vishwanathan2010graph}. However, we will not discuss them but primarily focus on the deep learning-based NRL model.

\subsubsection{GGS-NNs}
\begin{figure}[]
  \centering
  \includegraphics[width=3.5in,height=0.7in]{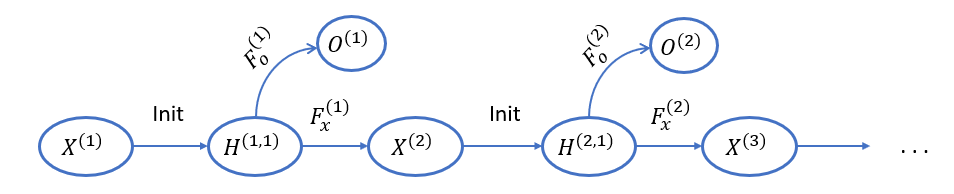}
  \caption{GGS-NNs architecture.}
  \label{GGS-NNs}
\end{figure}

To deal with graph-structured data, Li et al.~\cite{li2015gated} proposed a novel graph-based neural network model called Gated Graph Sequence Neural Networks (GGS-NNs) by extending the previous related GNNs model~\cite{scarselli2009graph}. Being different from the feature learning algorithm GNNs, the modification specially used gated recurrent units~\cite{cho2014learning} and modern optimization techniques. In addition, GGS-NN can produce sequence outputs, e.g., paths on a graph, rather than a single output. The basic component of GGS-NNs is GG-NNs containing three main parts: (1) node annotation process initializes node representation; (2) propagation model computes node representation of each of them; (3) output model, i.e., the model $\mathbf{o}_v = g(\mathrm{h}_v^{(T)}x_v)$ maps node representations with their labels to outputs, where the notations $h_v$ and $x_v$ denote a representation of node and node label, respectively. In graph level outputs, a graph level representation vector is defined as:
\begin{equation}
\mathrm{h}_g = \mathrm{tanh}(\sum_{v\in V} \sigma (\mathrm{i}(\mathrm{h}_v^{(T)},x_v))\odot
\mathrm{tanh}(\mathrm{j}(\mathrm{h}_v^{(T)},x_v))),
\end{equation}
where $\sigma(*,*)$ works as a soft attention mechanism by deciding the relevant nodes to the current graph level task. $\mathrm{i}(*,*)$ and $\mathrm{j}(*,*)$ are neural networks for computing inputs $h_v,x_v$ to real-valued vectors. The last two parts are the core processes, which map the graph to the output. As mentioned above, GGS-NNs can produce sequence outputs, e.g., $o^{(1)},o^{(2)},\ldots,o^{(K)}$, that is different from most graph representation learning algorithms. The architecture shown in Fig.~\ref{GGS-NNs} contains several GG-NNs operating in sequence to produce sequence outputs.

There are some new GNN-based models with innovations of involving new pooling strategies for graph-level representation. For example, Ying et al.~\cite{ying2018hierarchical} proposed a differentiable pooling strategy, which can learn hierarchical representations of graphs but it has the drawback of high computational complexity; Zhang et al.~\cite{zhang2019hierarchical} also designed a hierarchical graph pooling method without parameter. The proposed method has good performance of keeping key substructures of graph. Lee et al.~\cite{lee2019self} designed a pooling method on graph with involving self attention mechanism. This mechanism could pilot model focusing on important features like a human.

\subsubsection{CNN-Graphs-FLSF}
Generalizing CNN to graph-structured data is always a great challenge. To overcome this problem,  a novel convolutional neural network model was combined with fast localized spectral filtering for graphs (CNNs-Graphs) in~\cite{defferrard2016convolutional}. CNN is a powerful tool in images, video tasks, and natural language processing. There are several contributions to generalize the classical CNN from low dimensional data to high dimensional irregular domains, e.g., social networks and brain connectomes. At first, a novel convolutional filter named fast localized spectral filter on graphs was proposed by enhancing the previous GCNs algorithm~\cite{bruna2013spectral}. The fast localized spectral filter is a spectral approach based on spectral graph theory~\cite{chung1997spectral}. In general, the convolutional filter can be applied to recognize identical features of data. Spectral approaches are used here to offer a well-defined localization operator on graphs in spectral domain~\cite{shuman2013emerging}. Here, the graph Laplacian matrix~\cite{chung1997spectral} is leveraged as an essential operator in spectral graph analysis, and the formula definition has been given in Section 2 Definition 3.

In order to improve the computational efficiency of the CNN-Graphs model, the polynomial filter is first utilized to reduce the learning complexity, which can achieve the same complexity as classical CNN. In addition, to further reduce the model complexity, Chebyshev expansion~\cite{hammond2011wavelets} was utilized to overcome the high computational cost caused by the multiplication with the Fourier basis. Second, Graph coarsening is a necessary operation because pooling operation needs a valid group of data (they are neighborhoods for a graph) and coarsening phase of Graclus multilevel clustering algorithm~\cite{dhillon2007weighted} was employed to group similar vertices in graphs. Last, an efficient and fast pooling strategy was proposed by constructing a balanced binary tree and then applied pooling operation on the rearranged
vertices of the graph.

\section{Applications}
There are many NRL algorithms proposed based on deep learning models. In
practice, NRL and network embedding are commonly used in graph analytic tasks, such as node classification,
link prediction, clustering, and graph visualization. In this section, we will
discuss the applications of these NRL algorithms in the following.

\subsection{Node Classification}
Classification refers to dividing items into different categories. In network science, the most common applications in the graph analysis
task include node classification~\cite{Pan2016Tri} and graph classification. In addition, node or graph classification is often used as a
benchmark to evaluate the performance of node embedding. In the node classification application, labels are able to indicate different information, such as interest and affiliations. However, due to the limited amount of labeled data in reality and only a few available labels, semi-supervised learning is often considered to enhance the performance of the node classification task~\cite{kipf2016semi}~\cite{Hu2017Label}, or text classification~\cite{yang2016revisiting}. The learned latent representation is a real-value vector, which is convenient to be combined with traditional classification algorithm to address the problems, such as multi-label classification~\cite{grover2016node2vec}~\cite{perozzi2014deepwalk}~\cite{bui2017neural}. Besides, Wang et al.~\cite{wang2016structural} proposed a semi-supervised deep embedding algorithm, which can be used for link prediction and multi-label classification in blog-catalog networks. In heterogeneous networks, e.g., social networks, various types of network data take a great challenge for mining network data. Most representation learning algorithms just consider network structure. However, network vertices contain rich text information, which can be incorporated with NRL through matrix factorization so as to achieve high classification accuracy~\cite{yang2015network}. Another primary application of graph analysis task is graph classification, which is assigning graphs to several categories, such as classifying proteins based on biological function~\cite{hamilton2017inductive}~\cite{niepert2016learning}~\cite{yanardag2015deep}.

In order to present the general process of node classification, and give comparisons of performance with some representative methods, we briefly conduct a node classification on wiki dataset\footnote{https://linqs.soe.ucsc.edu/data}. The dataset is a webpage network consisting of 2,405 nodes, 17,981 links, and 20 classes. We utilize an open-source toolkit\footnote{https://github.com/thunlp/OpenNE}, which packs common methods and provides flexible parameters to control algorithms. Here, we evaluate performance with metrics of time consumption, Micro-F1 score, and Macro-F1 score.  The ratio of training is $0.5$. The results are shown in Table~\ref{tab:f1score}. As shown in the table, network embedding can be used to node classification. Different methods have different performance in terms of computational efficiency and embedding accuracy.

% Table generated by Excel2LaTeX from sheet 'Sheet1'
\begin{table}[htbp]
  \centering
  \caption{Node classification of wiki dataset by different network embedding algorithms}
    \begin{tabular}{|c|c|c|c|}
    \hline
    \textbf{Methods} & \textbf{Time} & \textbf{Micro-F1} & \textbf{Macro-F1} \\
    \hline
    LE~\cite{belkin2003laplacian}  & 1.13s & 0.36 & 0.15 \\
    \hline
    Node2vec~\cite{grover2016node2vec} & 43.34s & 0.66 & 0.53 \\
    \hline
    LINE~\cite{tang2015line} & 342s & 0.39 & 0.28 \\
    \hline
    SDNE~\cite{wang2016structural} &   2271s  &  0.63   & 0.51 \\
    \hline
    HOPE~\cite{Ou2016Asymmetric} & 1.62s & 0.60 & 0.43 \\
    \hline
    \end{tabular}%
  \label{tab:f1score}%
\end{table}%

\subsection{Link Prediction}
Another popular application of node embedding is link prediction by finding explicit or implicit links between nodes in graphs, for example, mining the relationship links in social networks~\cite{tang2015line}~\cite{tu2017transnet}~\cite{wang2017shifu}~\cite{Ou2016Asymmetric}. Link prediction is widely used to predict unknown interactions between nodes to observe links and properties. In addition, it can be used to recommend items via establishing links between users, such as predicting affinities between users and movies~\cite{berg2017graph}. Moreover, node embedding transforms a node into real-value vector, and similar node vectors tend to stay close in the latent learning space. This intrinsic characteristic helps a lot in the prediction of missing edges~\cite{grover2016node2vec}~\cite{wang2016structural} as close nodes are likely to have connections in the future. Recently, knowledge graph is becoming a hot topic in the network science domain, in which predicting missing relations of entities attracts lots of attention~\cite{nickel2016review}. Predicting unknown interactions between proteins in computational biology is a fundamental problem, which can be treated as the link prediction problem in graphs.

\subsection{Clustering}
Clustering is a traditional problem of machine learning. Graph clustering refers to nodes or graphs having similar affiliations or interests which are densely grouped together in a cluster. There are numerous applications of node clustering for community detection~\cite{fortunato2010community}, text categorization~\cite{defferrard2016convolutional,cao2016deep}, computational biology, and recommender systems~\cite{shi2019heterogeneous}, etc. Because nodes after embedding process are real-valued vectors, density-based clustering is able to leverage the vector to perform node clustering tasks in graphs~\cite{cao2015grarep,Wang2016Paired}. Furthermore, graph clustering is a powerful tool for chemical analysis. For example, it can be used to divide certain wines based on its chemical analysis information from different categories~\cite{tian2014learning}. Similar to classification, clustering also can be utilized to evaluate the performance of representation learning. In NRL algorithms~\cite{tang2015line}, the empirical experiment showed that it has achieved high performance compared with the methods by using the metric of matching authors to the belonging communities.

\subsection{Graph Visualization}
Graph visualization is a great way to help human understand and analyze sophisticated networks. The basic form of graph visualization is to project high dimensional network data into a 2D picture, where the same group nodes have the same color, and different node categories can be easily distinguished. For example, LINE~\cite{tang2015line} is able to visualize the same group authors in the same field, and the data come from co-authorship networks. There are several benefits of graph visualization. When a graph is visualized as a 2D image, it will be easy to reveal the real intrinsic structure of graphs, such as discovering the hidden structure or finding communities. In real applications, graph visualization has many varieties of applications through social science~\cite{chamberlain2017neural}~\cite{Freeman2000Visualizing} and biology visualization~\cite{niepert2016learning}. Furthermore, similar nodes stay close to each other in the 2D visualization figure as well as node clustering. Several researchers utilized graph visualization to present document categories in the visual form~\cite{wang2016structural}~\cite{Pan2016Tri}. Because node embedding associates with real-value vector, it often connects with dimension-reduced methods, such as PCA and t-SNE~\cite{Laurens2008Visualizing}, or other traditional methods. 

\begin{figure*}[]
  \centering
  \includegraphics[width=6.0in,height=1.5in]{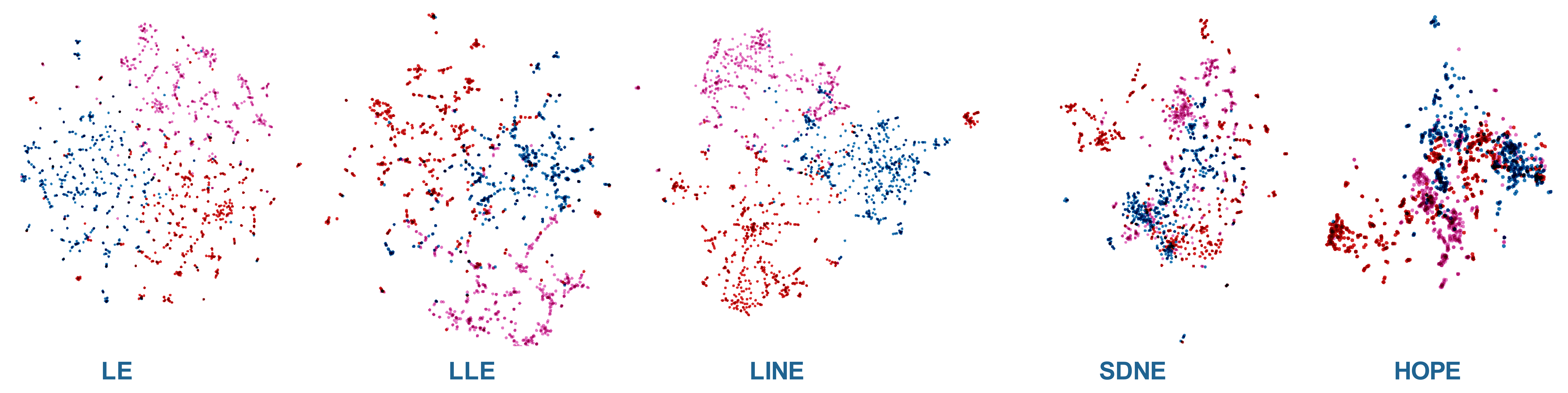}
  \caption{Graph visualization of 20Newsgroups dataset.}
  \label{20Newsgroups}
\end{figure*}

We visualize 20Newsgroups dataset\footnote{http://qwone.com/~jason/20Newsgroups/} using different NRL methods, i.e., LE~\cite{belkin2003laplacian}, LLE~\cite{roweis2000nonlinear}, LINE~\cite{tang2015line}, SDNE~\cite{wang2016structural}, and HOPE~\cite{Ou2016Asymmetric}. Different colors of nodes in pictures represent different classes. The results are shown in Fig~\ref{20Newsgroups}. We can see that nodes with the same colors are clustering together. That means network embedding could keep the original structural information of the graph.

\subsection{Other Applications}
Besides the general applications mentioned above, there are still a number of specific applications. Herein, we briefly give some examples in the following.

GGS-NNs~\cite{li2015gated} is a feature learning technique for graph-structured data. The novel characteristic of the algorithm is that it can produce a sequence of outputs rather than a single output. Different from the mentioned applications, this algorithm can be applied for BABI TASK, which contains 20 testing basic forms of reasoning tasks, such as deduction, induction, counting, and path-finding, etc. Duvenaud et al.~\cite{duvenaud2015convolutional} extended the application of NRL to predict the properties of new molecules. The algorithm is proposed based on circular fingerprints and has a better predictive performance on a series of tasks. Predicting protein function is an interesting topic of bioinformatics, such as OhmNet~\cite{zitnik2017predicting} leveraging multi-layers tissue networks to predict multicellular function, which achieved remarkable prediction accuracy. NRL is also an important way to detect community. For example, Li et al.~\cite{li2018community} proposed a novel embedding based method for community detection leveraging both attributes and structure information of graphs, Tu et al.~\cite{tu2018unified} proposed unified framework for community detection considering NRL and text modeling.

\section{Open issues}
Even though NRL is a powerful and general technique for graph analysis. There still remain lots of concrete open research problems.

\textbf{Large-scale graphs modeling.} According to our knowledge, few NRL algorithms can better handle large-scale networks, and most of them are just suitable for small-scale networks rather than that having hundreds of millions of nodes and links. With large-scale networks, researchers usually care about computational efficiency and how to combine heterogeneous network structure and multi-type node information. In addition, with the scale of graph increasing, the reconstruction vectors of graphs are becoming vague or inaccurate. That is a big problem when dealing with classification of large-scale graphs, and we believe that how to deal with large-scale network embedding still needs further exploration within the domain of NRL in the future.

\textbf{Model depth.} Deep learning has achieved a great performance improvement for image classification, handwriting recognition, etc., and has been widely applied for network representation learning in recent years, such as Graph Neural Networks, Graph Convolutional Networks, and Graph Autoencoders etc. Although there are many deep learning-based methods proposed, most of them are shallow models. Too many layers could cause an over-smooth problem and could not fully extract features of graphs. We still lack efficient network embedding approaches to cope with these problems. There are some NRL works adopting ideas from deep learning models, such as DenseNet, graph convolution of different scales. Deep graph learning is still an open issue for researchers to further study.

\textbf{Interpretability.} We know that many deep learning models behave as black-boxes, which causes the problem of interpretability. However, lots of graph learning methods derived from deep learning methods need interpretability. For example, graph learning methods are for recommend system or decision-making system. There are few researchers to address this problem with graph learning. So, interpretability is crucial and even more challenging as complex characteristics of graph data.

\textbf{Robustness training and adversarial attacks.} Most NRL algorithms rely on ideal graph-structured data. However, in most cases, data are often truncated, missing, fuzzy, lopsided and we cannot get ideal information. Few papers are discussing how to well handle robustness with deep learning-based NRL methods. Even though some works involve GAN mechanism to enhance the robustness of embedding, these algorithms are inefficient and only suitable for specific networks. In addition, deep learning model is sensitive to adversarial attacks. So, these deep learning-based NRL methods are inherently unable to overcome the attack problem. In summary, the robust graph learning techniques still need to be further explored.

\section{Conclusion}\label{section:conclusion}
In this paper, we review the NRL algorithms including TFL models and deep learning-based models. These NRL algorithms can learn to reconstruct the representation of graphs. We first give a brief introduction about TFL and then separately discuss the NRL algorithms by focusing on the graph sources (edges and node attributes). There are too many categories of NRL, and we mainly pay attention to these deep learning-based models. We propose three taxonomies of graphs embedding from the deep learning perspective as shown in Table~\ref{algorithms_category}. The most important contributions are the parts that introduce NRL algorithms. Also, we summarize the application of graphs embedding in the aspects of classification and semi-supervised learning, link prediction, clustering, etc. Finally, we emphasize that NRL has great promising future research directions in the field of network science.

\bibliographystyle{IEEEtran}
\bibliography{IEEEabrv,references}

\EOD

\end{document}